\documentclass[
    ,final            
  ]
  {aipproc}

\layoutstyle{8x11single}
\usepackage{amsmath,amssymb,amsopn,amsfonts,bm,empheq,amscd,
accents,framed,delarray,epic}

\newcommand{\etal}{{\em et al.}}
\newcommand{\ph}{\phantom}

\begin{document}

\title{Status of Precision Extractions of $\alpha_s$ and Heavy Quark Masses}


\author{Jens Erler}{
address={Departamento de F\'isica Te\'orica, Instituto de F\'isica, \\
Universidad Nacional Aut\'onoma de M\'exico, 
04510 M\'exico D.F., M\'exico}
}

\begin{abstract}
An overview of precision determinations of the strong coupling constant, as well as the top, bottom and charm quark masses is presented.
\end{abstract}

\maketitle


\section{Introduction}
Precise knowledge of the fundamental parameters of the Standard Model (SM) of the strong and electroweak interactions 
is a key goal in its own right, and will also allow to make theoretical predictions of physical processes in which they enter,
which in turn permits to test the internal consistency of our basic theoretical framework. 
A recent example is the determination of Higgs boson branching ratios, 
which opens the window to precision physics of the electroweak symmetry breaking sector.
They strongly depend on the values of the heavy quark masses whose uncertainties could become
the limiting factor in the future.
Due to its large mass, $m_t$, the top quark is generally believed to play a special role here.
For example, whether we live in a stable or meta-stable vacuum of the SM, or whether new physics is needed
to provide a sufficiently long-lived ground state, depends crucially on the value of $m_t$.

Furthermore, the fundamental input parameters of the SM may be related in the context of new physics underlying it,
by virtue of new symmetries or other unifying principles yet to be discovered.  
The prime example is the observed approximate unification of gauge couplings in the minimal supersymmetric 
extension of the SM.
This may be one of the very rare possibilities to access near Planck scale physics.
High precision is mandatory for this, so that further progress requires specifically 
higher precision in the strong coupling constant, $\alpha_s$,
which is by far the least known of the three gauge couplings.
Similarly, the third generation fermion masses may obey relatively simple relations,
calling for precision extractions of the bottom quark mass, $m_b$, alongside with $m_t$.

We will first address $m_t$ determinations and the various sources of uncertainties, 
both experimental and theoretical. 
This will be confronted with indirect, {\em i.e.\/}, electroweak constraints. 
Then we turn to recent results on $m_b$ which we discuss together with the charm quark mass, $m_c$,
because the most precise approaches tend to apply to both, although the uncertainties may be quite different
(we will not cover methods to extract their ratio, $m_b/m_c$, directly).
Finally, we review the status of $\alpha_s$.
Some more space will be dedicated to $Z$ decay extractions, 
as the only ones which are not actually dominated by the theory error.
We also enter more into $\tau$ decays as the alternative electroweak process constraining $\alpha_s$ 
and where the most pressing issue is perturbative rather than non-perturbative QCD.

\section{Top Quark Mass}
\subsection{Hadron collider combination}
\begin{figure}
\label{figmt}
  \includegraphics[height=.5\textheight]{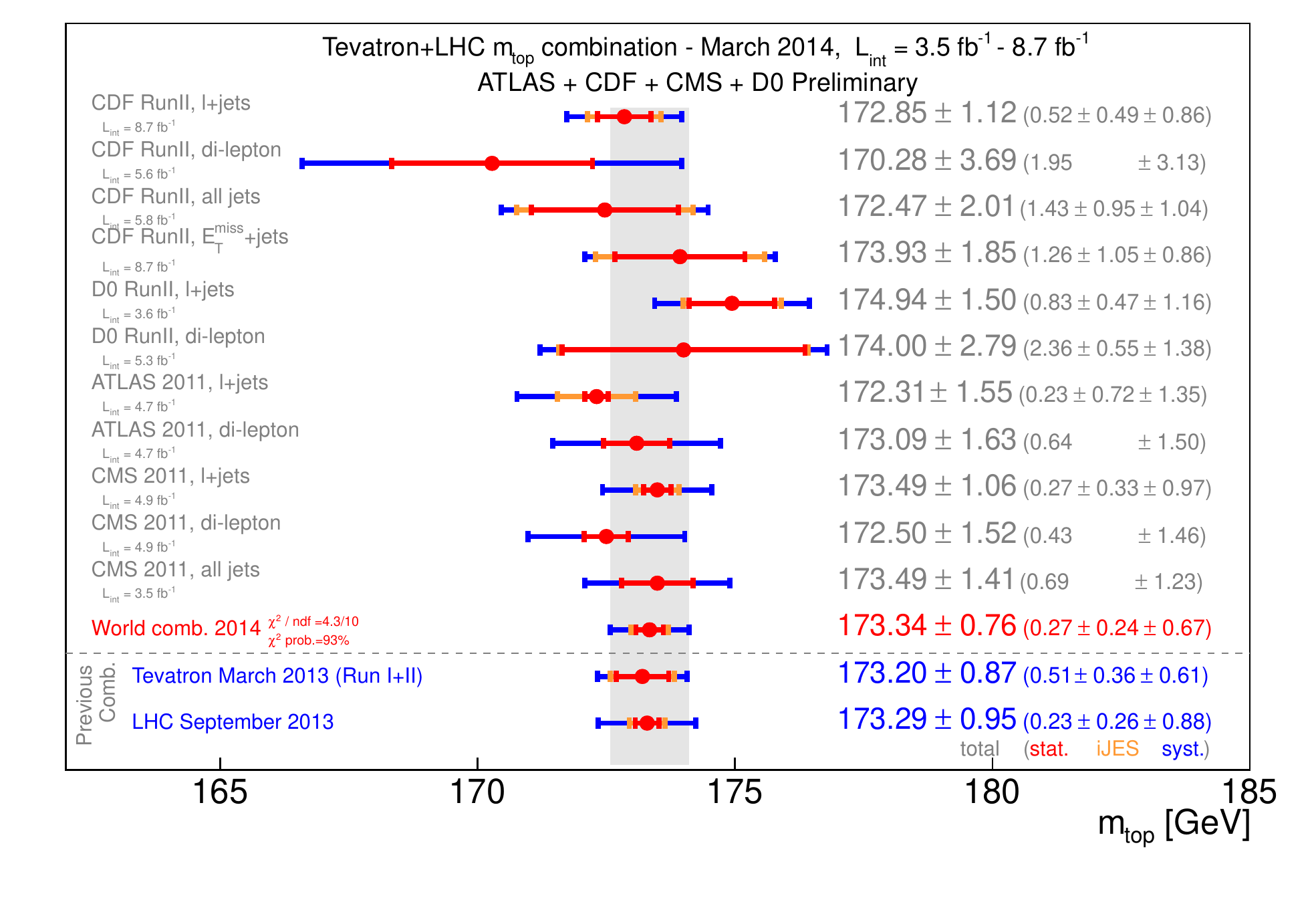}
  \caption{Top quark (pole) mass measurements by the CDF, D\O\, ATLAS and CMS Collaborations, 
  and their combination~\cite{ATLAS:2014wva}. 
  The CDF entry denoted $E_\gamma^{\rm miss}$ + jets are decays involving taus or unidentified muons or electrons.
  The ATLAS and CMS results are exclusively from LHC runs at the center of mass energy of 7~TeV.
  Notice, that the value of $\chi^2 = 4.3$ for 10~effective degrees of freedom is excellent, if not somewhat low.
  Also shown are the most recent combinations internal to the Tevatron and the LHC, respectively, 
  but note that the input measurements are not identical to those used for the global average.
  (Figure reprinted from Reference~\cite{ATLAS:2014wva}.)}
\end{figure}

Recently, the CDF and D\O\ Collaborations at the Tevatron (Fermilab) and the ATLAS and CMS Collaborations at the LHC (CERN)
performed the first common hadron collider average of $m_t$ measurements~\cite{ATLAS:2014wva}.
The individual input measurements and the combination are shown in Figure~\ref{figmt},
represented as values of the top quark pole mass, $m_t^{\rm pole}$, 
because all of these results rely on the kinematic reconstruction of top quark decays.
It is instructive to consider the breakdown of the experimental uncertainty,
which --- including an additional QCD uncertainty as described below --- can be written as
\begin{eqnarray}
m_t^{\rm pole} &=&
173.34 \pm 0.27_{\rm stat} \pm 0.33_{\rm JSF} \pm 0.25_{\rm bJES} \pm 0.54_{\rm theroy\; \&\; model} \pm 0.2_{\rm other} 
\pm 0.5_{\rm QCD} \mbox{ GeV} \nonumber \\
&=& 173.34 \pm 0.76_{\rm exp} \pm 0.5_{\rm QCD} \mbox{ GeV} \nonumber \\
&=& 173.34 \pm 0.91_{\rm total} \mbox{ GeV}.
\label{eqmt}
\end{eqnarray}
Some of these errors are not or only weakly correlated between the various channels and experiments,
including the one from the jet energy scale factor (JSF) which is determined for each detector and enters differently across channels, 
and, of course, the statistical component (stat).
On the hand, the uncertainty originating from the $b$ jet energy scale (bJES) is strongly correlated even for different detectors
because it is estimated using common theory, assumptions and inputs. 
The same applies to the theory \& model components, which include error sources associated with radiative corrections, 
color reconnection, and parton distribution functions (PDFs), all of which clearly affect different experiments in similar ways. 
Another model error arises from the background Monte Carlo generators,
but this induces correlations only between the same analysis channels.

The kinematic mass determination should --- in the absence of confinement --- correspond to the pole mass of the quark.
However, the exact relation between the kinematic fit variable (called hereafter the Monte Carlo mass, $m_t^{\rm MC}$)
and $m_t^{\rm pole}$ is not known and they may well differ by an amount of the order of the strong interaction scale.
Moreover, $m_t^{\rm pole}$ is not the quantity that actually enters the theoretical expressions of electroweak observables,
such as for the mass of the $W$ boson, the weak mixing angle, or Higgs boson production and decay amplitudes.
Rather, these formul\ae\ are more appropriately written in terms of a short-distance mass definition
--- such as the $\overline{\rm MS}$ mass, $\overline{m}_t$, defined within the modified minimal subtraction regularization 
and renormalization scheme ---
so as to avoid renormalon~\cite{Beneke:1998ui} ambiguities associated with long-distance mass definitions like $m_t^{\rm pole}$
and the corresponding deterioration of the perturbative series. 
The renormalon ambiguity reappears in the relation between $m_t^{\rm pole}$ and $\overline{m}_t$,
but this will then be the only place.
Thus, one can conveniently estimate the renormalon uncertainty as the ${\cal O}(\alpha_s^3)$ term 
in the conversion formula~\cite{Chetyrkin:1999qi,Melnikov:2000qh} between $m_t^{\rm pole}$ and $\overline{m}_t$ 
which amounts to about 0.5~GeV.
Assuming that $m_t^{\rm MC}$ does not differ by more than this from $m_t^{\rm pole}$ 
one may finally combine this QCD error with the experimental uncertainty as done in Equation~\eqref{eqmt}.
With this assumption we can summarize these considerations by writing,
\begin{equation}
\label{mtMC}
m_t^{\rm MC} \simeq m_t^{\rm pole} = \overline{m}_t(\overline{m}_t) + 9.65 \pm 0.50 \mbox{ GeV}.
\end{equation}
Note, however, that $m_t^{\rm MC}$ itself does not carry a renormalon ambiguity,
and so it can be avoided by bypassing $m_t^{\rm pole}$ which only appears as an intermediary. 
Specifically, Hoang and Stewart~\cite{Hoang:2008xm} relate $m_t^{\rm MC}$ to
the ``MSR-mass'', $m_t^{\rm MSR}$~\cite{Hoang:2008yj}, evaluated at a rather low scale.
As shown in Reference~\cite{Hoang:2008yj} this amounts to
\begin{equation}
\label{mtMSR}
m_t^{\rm MC} = m_t^{\rm MSR}(3^{+6}_{-2} \mbox{ GeV}) = \overline{m}_t(\overline{m}_t) + 9.6^{+0.6}_{-0.3} \mbox{ GeV},
\end{equation}
which in practice does not differ by much from Equation~\eqref{mtMC}, but is on a more solid theoretical footing.

\begin{figure}
\label{figmt2}
  \includegraphics[height=.5\textheight]{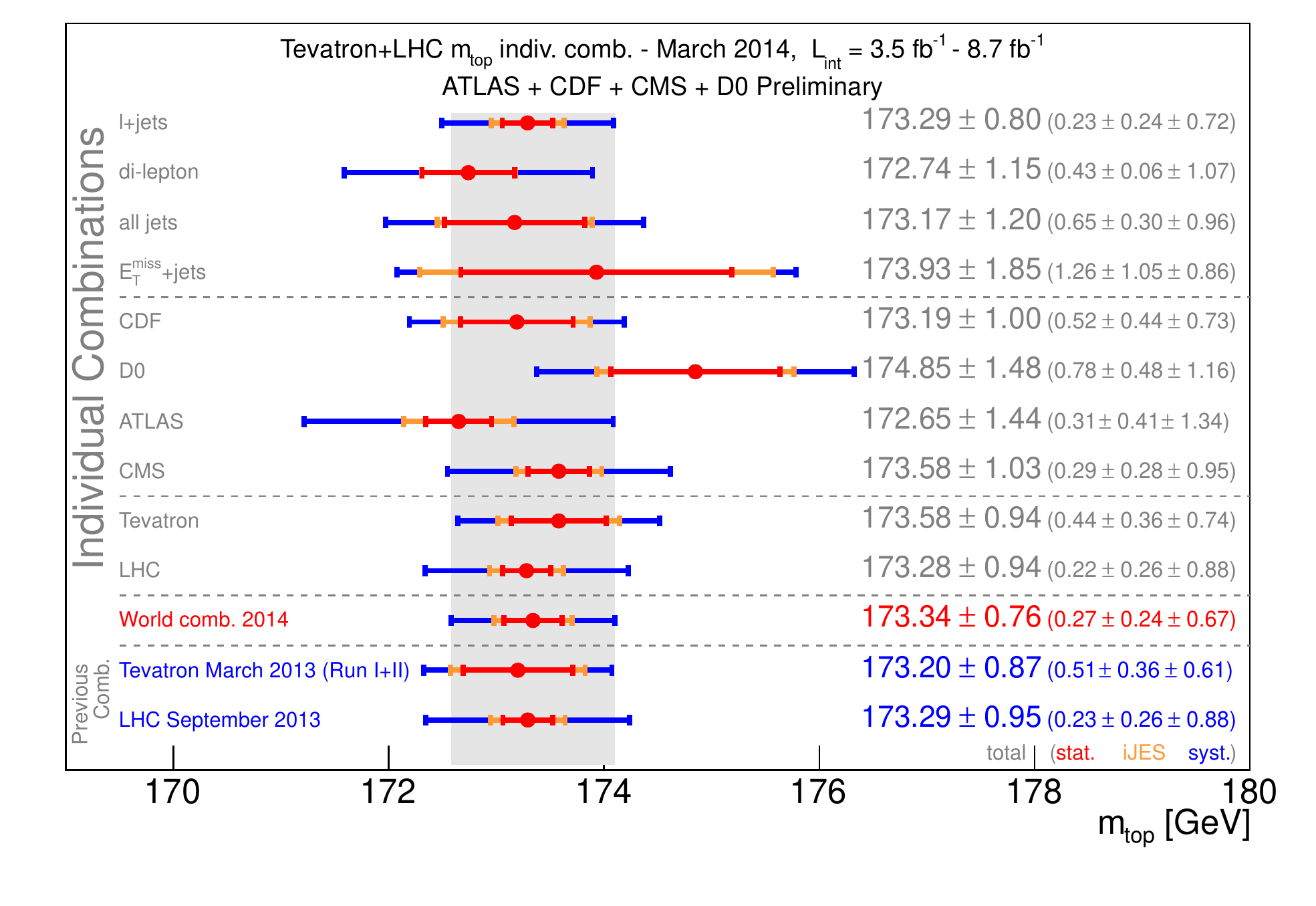}
  \caption{Breakdown of the top quark mass combination by the CDF, D\O\, ATLAS and CMS Collaborations~\cite{ATLAS:2014wva}
  according to decay channel, detector, and collider, respectively.
  (Figure reprinted from Reference~\cite{ATLAS:2014wva}.)}
\end{figure}

Figure~\ref{figmt2} shows partial $m_t$ combinations.
The lepton + jets channel is the most precise as it combines the advantages of clean lepton detection
and good statistics.  
The all jets channel is, of course, more difficult in a hadron environment, while the reconstruction in the di-lepton channel
is complicated by two escaping neutrinos.
Very good agreement is observed.
Notice, that the Tevatron combination in Figure~\ref{figmt2} has a larger error,
than the previous Tevatron average~\cite{CDF:2013jga} of March 2013.
This is because the Tevatron/LHC combination does not actually use all the available experimental information,
but rather employs (for simplicity) only the most precise measurements per experiment and channel.

\subsection{After the combination}
This very consistent picture changed after new results became available following the combination.
In Reference~\cite{Abazov:2014dpa} the D\O\ Collaboration presented a new result from the lepton + jets final state,
\begin{equation}
\label{eqmtD0}
m_t^{\rm pole}  = 174.98 \pm 0.58_{\rm stat\; \&\; JSF}  \pm 0.49_{\rm syst} \mbox{ GeV} =
174.98 \pm 0.76_{\rm exp} \mbox{ GeV}
\qquad\qquad(\mbox{D\O\ lepton + jets 2014)},
\end{equation}
which constitutes the most precise single measurement and matches the world average in Equation~\eqref{eqmt} in precision.
While this agrees with the previous D\O\ result in the same channel shown in Figure~\ref{figmt}, 
it is significantly higher than the Tevatron/LHC combination.
Subsequently, the Tevatron groups performed a new average~\cite{Tevatron:2014cka},
\begin{equation}
\label{eqmtTevatron}
m_t^{\rm pole}  = 174.34 \pm 0.37_{\rm stat}  \pm 0.52_{\rm syst} \mbox{ GeV} =
174.34 \pm 0.64_{\rm exp} \mbox{ GeV}
\qquad\qquad(\mbox{Tevatron 2014)}.
\end{equation}
Very recently, the ATLAS Collaboration published its first all-jet analysis~\cite{Aad:2014zea},
\begin{equation}
\label{eqmtATLAS}
m_t^{\rm pole}  = 175.1 \pm 1.4_{\rm stat}  \pm 1.2_{\rm syst} \mbox{ GeV} =
175.1 \pm 1.8_{\rm exp} \mbox{ GeV}
\qquad\qquad(\mbox{ATLAS all jets)},
\end{equation}
which is high but consistent within errors.
On the other hand, the first result from the LHC running at the center of mass energy of 8 TeV yields the low value~\cite{CMS:2014ima},
\begin{equation}
\label{eqmtCMS}
m_t^{\rm pole}  = 172.04 \pm 0.19_{\rm stat\; \&\; JSF}  \pm 0.75_{\rm syst} \mbox{ GeV} =
172.04 \pm 0.77_{\rm exp} \mbox{ GeV}
\qquad\qquad(\mbox{CMS lepton + jets, 8 TeV)}.
\end{equation}
Combined with previously published measurements by CMS, 
this gives $m_t^{\rm pole}  = 172.22 \pm 0.73 \mbox{ GeV}$~\cite{CMS:2014ima}
which is $2.2~\sigma$ lower than the most recent Tevatron combination in Equation~\eqref{eqmtTevatron}.
Furthermore, the two most precise individual determinations in Equations~\eqref{eqmtD0} and \eqref{eqmtCMS}, 
both from the lepton + jets final state, are in conflict with each other by $2.7~\sigma$ --- or possibly more in case of correlated systematics. 
CMS also released an all-jets result from the 8 TeV run~\cite{CMS:2014rta},
\begin{equation}
\label{eqmtCMSjets}
m_t^{\rm pole} = 172.08 \pm 0.36_{\rm stat\; \&\; JSF}  \pm 0.83_{\rm syst} \mbox{ GeV} =
172.08 \pm 0.90_{\rm exp} \mbox{ GeV}
\qquad\qquad(\mbox{CMS all jets, 8 TeV)},
\end{equation}
which is in perfect agreement with Equation~\eqref{eqmtCMS} and by far the most precise determination in this channel.
It is even significantly more precise than the all jets combination in Figure~\ref{figmt2}.

Looking ahead, Reference~\cite{CMS:2013wfa} gives projections for future kinematic mass determinations at CMS, namely
\begin{eqnarray}
\int {\cal L} dt = 30 \mbox{ fb}^{-1} \mbox{ at } \sqrt{s} = 13 \mbox{ TeV}: &
\Delta m_t = \pm 0.15_{\rm stat} \pm 0.60_{\rm syst} \mbox{ GeV} = \pm 0.62_{\rm exp} \mbox{ GeV},  \\
\int {\cal L} dt = 300 \mbox{ fb}^{-1} \mbox{ at } \sqrt{s} = 14 \mbox{ TeV}: &
\Delta m_t = \pm 0.05_{\rm stat} \pm 0.44_{\rm syst} \mbox{ GeV} = \pm 0.44_{\rm exp} \mbox{ GeV},  \\
\int {\cal L} dt = 3000 \mbox{ fb}^{-1} \mbox{ at } \sqrt{s} = 14 \mbox{ TeV}: &
\Delta m_t = \pm 0.01_{\rm stat} \pm 0.20_{\rm syst} \mbox{ GeV} = \pm 0.20_{\rm exp} \mbox{ GeV}.
\end{eqnarray}
The conversion error, as illustrated for example in Equation~\eqref{mtMSR} may also improve with more data.

\subsection{Alternative methods}
Kinematic mass determinations have reached a precision at which the QCD error displayed explicitly in Equation~\eqref{eqmt}
becomes significant and it will eventually dominate the overall uncertainty. 
To make progress and optimal use of the huge sample sizes from future LHC runs (especially in the high-luminosity phase)
it is mandatory to develop alternative methods, including in particular those allowing to directly extract a short-distance mass.

One possibility is to derive $m_t$ from the inclusive $t\bar{t}$ production cross section, $\sigma_{t\bar t}$.
ATLAS~\cite{Aad:2014kva} measured $\sigma_{t\bar t}$ using both 7~TeV and 8~TeV data and
obtained $m_t^{\rm pole} = 172.9^{+2.5}_{-2.6}$~GeV. 
This is consistent within errors with the results from kinematic reconstruction, but there is a $1.7~\sigma$ tension 
between the values derived from the 7~TeV and 8~TeV data sets. 
While ATLAS did not provide a direct determination of a short-distance mass.
the authors of Reference~\cite{Alekhin:2013nda} demonstrated the feasibility of such an approach 
by performing a fit to Tevatron and LHC $\sigma_{t\bar t}$ data and obtained,
\begin{equation}
\overline{m}_t(\overline{m}_t) = 162.3 \pm 2.3 \mbox{ GeV}
\qquad\qquad(\sigma_{t\bar t}).
\end{equation}
Reference~\cite{CMS:2013wfa} estimates that a future inclusive cross section determinations of $m_t$
to $\lesssim 1$~GeV precision is optimistic but at least conceivable.

An alternative observable suggested in Reference~\cite{Alioli:2013mxa} is the normalized differential distribution 
of the $t\bar{t} + 1$~jet cross section.
The authors state that this observable has the potential to provide $m_t$ to within $\pm 1$~GeV or better.

\subsection{Top quark mass from electroweak precision data}
\begin{figure}
\label{figmhmt}
  \includegraphics[height=.5\textheight]{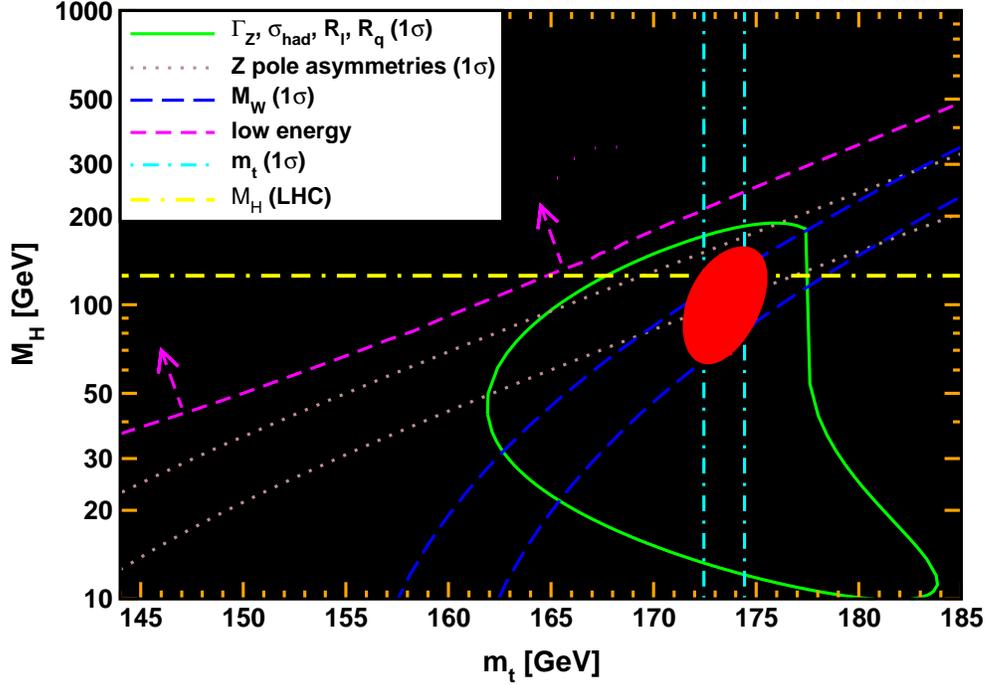}
  \caption{Fit result and one-standard-deviation (39.35\% for the closed contours and 68\% for the others) uncertainties in $M_H$ 
  as a function of $m_t$ for various inputs, and the 90\% CL region ($\Delta \chi^2 = 4.605$) allowed by all data.  
  $\alpha_s(M_Z) = 0.1185$ is assumed except for the fits including the $Z$~lineshape. 
  The width of the horizontal dashed-dotted (yellow) band is not visible on the scale of the plot.
  (Figure reprinted from Reference~\cite{Erler:2014}.)}
\end{figure}

One usually uses the $m_t$ measurements from the hadron colliders to constrain the Higgs boson mass, $M_H$, 
and compare the result with the reconstructed $M_H$ from the Higgs decays observed at the LHC~\cite{Aad:2012tfa,Chatrchyan:2012ufa}.
However, one can also reverse this and use $M_H$ from the LHC as input and predict $m_t$
from a global fit to all electroweak precision data excluding $m_t$.
This indirect prediction is for the $\overline{\rm MS}$ mass definition,
\begin{equation}
\overline{m}_t(\overline{m}_t) = 167.1 \pm 2.0 \mbox{ GeV}
\qquad\qquad(\mbox{global electroweak fit)},
\end{equation}
which can be converted in the end to the pole mass, $m_t^{\rm pole} = 177.0 \pm 2.1$~GeV. 
This is about $1.6~\sigma$ higher than the value in Equation~\eqref{eqmt}, as a reflection of the fact that there is some tension
between the measurement of the $W$ boson mass, $M_W$, and the SM prediction.
This is illustrated in Figure~\ref{figmhmt}, where one can see 
that the long-dashed (blue) $M_W$ contour prefers higher $m_t$ or lower $M_H$ values.
Indeed, the $M_W$ world average is dominated by $M_W = 80.387 \pm 0.016$~GeV from the Tevatron~\cite{Aaltonen:2013iut}
which in turn is $1.5~\sigma$ high.

Note that the fits corresponding to Figure~\ref{figmhmt} were done before the hadron collider average~\cite{ATLAS:2014wva}
in Equation~\eqref{eqmt} became available. 
For the fits performed in Reference~\cite{Erler:2014} a simplified averaging procedure had been applied,
with the result $m_t = 173.24 \pm 0.81$~GeV, in excellent agreement with Equation~\eqref{eqmt} 
even though this was done about six months earlier and the employed data sets were not identical.

\section{Bottom and Charm Quark Masses}
\subsection{Recent bottom quark mass determinations}
It is desirable and indeed possible to determine the bottom quark mass to the percent level and better.
This level of accuracy rules out any approach in which the bottom quark pole mass is extracted,
as this would introduce a conversion error analogous and comparable in size to
the $\pm 0.5_{\rm QCD}$~GeV uncertainty in Equation~\eqref{eqmt} for the case of $m_t$.
Thus, modern analyses either extract the $\overline{\rm MS}$ mass definition, $\overline{m}_b(\overline{m}_b)$, directly,
or some other short-distance mass to be converted in the end to $\overline{m}_b(\overline{m}_b)$.

The most recent precision determinations of $\overline{m}_b(\overline{m}_b)$ are collected in Table~\ref{tab:mb}.
The uncertainties are often strongly dominated by theory, 
and with one exception the methods are based on either QCD sum rules or on lattice simulations.
The bottom quark is ideal for perturbative methods (including sum rules) since $\alpha_s(\overline{m}_b) \sim 0.23$ is relatively small,
while at the same time the additional scale introduced by $m_b$ presents a challenge for the lattice.
The mutual agreement between all determinations except for the Borel sum rule analysis is very good and consistent with
the weighted average,
\begin{equation}
\label{eqmb}
\overline{m}_b(\overline{m}_b) = 4175 \pm 5 \mbox{ MeV} 
\qquad\qquad(\mbox{weighted uncorrelated average)}.
\end{equation}
Of course, such a least-square average is problematic because the nominal uncertainties are theory dominated,
making it difficult to quantify them and to assess their distribution.
Even more importantly, while correlations are certainly present they are difficult to trace and have been ignored.
This tends to exaggerate the apparent consistency, to underestimate the overall error and may strongly bias the central value. 
In addition, not all the quoted errors are uncontroversial to begin with,
as some components may have been underestimated or there may be unknown or misjudged effects compromising the analyses. 
Thus, the precision indicated in Equation~\eqref{eqmb} should rather be interpreted as a hopefully realistic future goal.
For the time being, it seems fair to state that we now know $\overline{m}_b(\overline{m}_b)$ to approximately 0.5\% precision 
or about 20~MeV. 

\begin{table}
\begin{tabular}{clllcc}
\hline
     \tablehead{1}{c}{b}{Result}
  & \tablehead{1}{c}{b}{Approach}
  & \tablehead{1}{c}{b}{Observable}
  & \tablehead{1}{c}{b}{Authors/Group}
  & \tablehead{1}{c}{b}{Reference} 
  & \tablehead{1}{c}{b}{arXiv.org} \\
\hline
$4196 \pm 23$~MeV & lattice ($n_f = 4$) & $\Gamma(\Upsilon, \Upsilon^\prime\to e^+e^-) $ & HPQCD & \cite{Colquhoun:2014ica} & 
1408.5768 \\
$4174 \pm 24$~MeV & lattice ($n_f = 4$) & pseudoscalar current & HPQCD & \cite{Chakraborty:2014aca} & 1408.4169 \\
$4201 \pm 43$~MeV & N$^3$LO PQCD & $M_\Upsilon$ & Ayala \etal & \cite{Ayala:2014yxa} & 1407.2128 \\
$4169 \pm \ph{9}9$~MeV & fiteenth moment sum rule & $\Upsilon$(1S--6S) & Penin and Zerf & \cite{Penin:2014zaa} & 1401.7035 \\
$4247 \pm 34$~MeV & Borel sum rule & $f_B$ and $f_{B_s}$ & Lucha \etal & \cite{Lucha:2013gta} & 1305.7099 \\
$4166 \pm 43$~MeV & lattice ($n_f = 2 + 1$) + PQCD & $M_\Upsilon$ and $M_{B_s}$ & HPQCD & \cite{Lee:2013mla} & 1302.3739 \\
$4235 \pm 55$~MeV & tenth moment sum rule & $\Upsilon$(1S--4S) and $R$ & Hoang \etal & \cite{Hoang:2012us} & 1209.0450 \\
$4171 \pm \ph{9}9$~MeV & pinched moment sum rule & $\Upsilon$(1S--4S) and $R$ & Bodenstein \etal & \cite{Bodenstein:2011fv} & 
1111.5742 \\
$4177 \pm 11$~MeV & Borel and $Q^2$-moments sum rule & $\Upsilon$(1S--6S) & Narison & \cite{Narison:2011rn} & 1105.5070 \\
$4180^{+50}_{-40} \ph{P}$~MeV & lattice + PQCD & $M_{\Upsilon(2S)}$ & Laschka \etal & \cite{Laschka:2011zr} & 1102.0945 \\
$4163 \pm 16$~MeV & second moment sum rule & $\Upsilon$(1S--4S) and $R$ & Chetyrkin \etal & \cite{Chetyrkin:2010ic} & 
1010.6157 \\
\hline
\end{tabular}
\caption{Most recent determinations of the bottom quark $\overline{\rm MS}$ mass, $\overline{m}_b^{(n_f = 5)}(\overline{m}_b)$,
defined in the five-flavor effective theory.
Only results with a quoted uncertainty of less than 60~MeV are included.
See body of text for details.}
\label{tab:mb}
\end{table}

While one aims to determine $\overline{m}_b(\overline{m}_b)$ as well as possible,
the accuracy in the na\"ive average in Equation~\eqref{eqmb} would be sufficient for most practical purposes
in the decades to come. 
Therefore, the intermediate-term future goal has to be to increase the transparency in the error estimates of individual determinations,
to derive a more appropriate combination procedure, 
and finally to reach a consensus that indeed the global average approaches the per mille level.

\subsection{Recent charm quark mass determinations}
Most of the general remarks regarding $\overline{m}_b(\overline{m}_b)$ carry over to $\overline{m}_c(\overline{m}_c)$,
for which the most recent precision determinations are shown in Table~\ref{tab:mc}.
Again most of the uncertainties are strongly dominated by theory, 
and in this case there are two alternatives to the sum rule and lattice QCD approaches.
The mutual agreement is seen to be even more impressive than in the $m_b$ case, 
with only the findings by the ETM Collaboration being somewhat high.
The weighted average is,
\begin{equation}
\label{eqmc}
\overline{m}_c(\overline{m}_c) = 1276 \pm 4 \mbox{ MeV} 
\qquad\qquad(\mbox{weighted uncorrelated average)}.
\end{equation}
This is clearly dominated by the very recent HPQCD simulation~\cite{Chakraborty:2014aca}. 
Note, that $m_c$ is generally a more natural target for the lattice since $\overline{m}_c(\overline{m}_c)$ is not much larger 
than the strong interaction scale, while perturbative methods suffer from the rather large $\alpha_s(\overline{m}_c)\approx 0.40$.
Thus, one way to interpret the table is to identify the value from Reference~\cite{Chakraborty:2014aca} 
as the world average and to consider the remaining results as corroborating support of its validity.

\begin{table}
\begin{tabular}{clllcc}
\hline
     \tablehead{1}{c}{b}{Result}
  & \tablehead{1}{c}{b}{Approach}
  & \tablehead{1}{c}{b}{Observable}
  & \tablehead{1}{c}{b}{Authors/Group}
  & \tablehead{1}{c}{b}{Reference} 
  & \tablehead{1}{c}{b}{arXiv.org} \\
\hline
$1275.8 \pm \ph{4}5.8$~MeV & lattice ($n_f = 4$) & pseudoscalar current & HPQCD & \cite{Chakraborty:2014aca} & 1408.4169 \\
$1348\ph{.8} \pm 46\ph{.8}$~MeV & lattice ($n_f = 2 + 1+ 1$) & $M_D$ and $M_{D_s}$ & ETM & \cite{Carrasco:2014cwa} & 
1403.4504 \\
$1274\ph{.8} \pm 36\ph{.8}$~MeV & lattice ($n_f = 2$)  & $M_D$ and $M_{D_s}$ & ALPHA & \cite{Heitger:2013oaa} & 1312.7693 \\
$1240\ph{.8}^{+50}_{-30} \ph{R.8}$~MeV & PDF + HT fit & DIS & Alekhin \etal & \cite{Alekhin:2013nda} & 1310.3059 \\
$1260\ph{.8} \pm 65\ph{.8}$~MeV & NLO fit & $c\bar{c}$ cross section & H1 and ZEUS & \cite{Abramowicz:1900rp} & 1211.1182 \\
$1262\ph{.8} \pm 17\ph{.8}$~MeV & $Q^2$-moments sum rule & $J/\Psi$, $\Psi$(2S--6S) & Narison & \cite{Narison:2011rn} & 
1105.5070 \\
$1260\ph{.8} \pm 36\ph{.8}$~MeV & lattice ($n_f = 2 + 1$) & $3 M_{J/\Psi} + M_{\eta_c}$ & PACS--CS & \cite{Namekawa:2011wt} &
1104.4600 \\
$1279\ph{.8} \pm \ph{9}9\ph{.8}$~MeV & pinched moment sum rule & $J/\Psi$, $\Psi^\prime$ and $R$ & Bodenstein \etal & 
\cite{Bodenstein:2011ma} & 1102.3835 \\
$1282\ph{.8} \pm 24\ph{.8}$~MeV & first moment sum rule & $J/\Psi$, $\Psi^\prime$ and $R$ & Dehnadi \etal & \cite{Dehnadi:2011gc} 
& 1102.2264 \\
$1280\ph{.8}^{+70}_{-60} \ph{R.8}$~MeV & lattice + PQCD & $M_{h_c(1P)}$ & Laschka \etal & \cite{Laschka:2011zr} & 1102.0945 \\
$1279\ph{.8} \pm 13\ph{.8}$~MeV & first moment sum rule & $J/\Psi$, $\Psi^\prime$ and $R$ & Chetyrkin \etal & 
\cite{Chetyrkin:2010ic} & 1010.6157 \\
\hline
\end{tabular}
\caption{Most recent determinations of the charm quark $\overline{\rm MS}$ mass, $\overline{m}_c^{(n_f = 4)}(\overline{m}_c)$,
defined in the four-flavor effective theory.
Only results with a quoted uncertainty not exceeding 70~MeV are included.
See body of text for details.}
\label{tab:mc}
\end{table}

\subsection{$m_b$ and $m_c$ from Lattice Gauge Theory}
We now discuss the various determinations of $m_b$ and $m_c$ in slightly more detail, 
starting with lattice simulations.

The HPQCD Collaboration~\cite{Colquhoun:2014ica,Chakraborty:2014aca,Lee:2013mla} announced three rather recent
results on $m_b$, two of which using $n_f = 4$ dynamical quarks.
In Reference~\cite{Colquhoun:2014ica} the collaboration analyzes the decay rates $\Gamma (\Upsilon \to e^+ e^-)$ 
and $\Gamma (\Upsilon' \to e^+ e^-)$ and moments of the vector-current correlator. 
The $b$ quarks are treated in non-relativistic QCD (NRQCD) while sea quarks are
included following the highly improved staggered quark (HISQ) formalism~\cite{Follana:2006rc}.
Similarly, Reference~\cite{Chakraborty:2014aca} studies moments of the pseudoscalar correlator.
In addition to $m_b$, values for $m_c$, $\alpha_s$, $m_b/m_c$, and $m_c/m_s$ are computed.
Reference~\cite{Lee:2013mla} presents a determination of $m_b$ to ${\cal O}(\alpha_s^2)$ in lattice perturbation theory,
including partial contributions at  ${\cal O}(\alpha_s^3)$. 
Non-perturbative input comes from the calculation of the $\Upsilon$ and $B_s$ energies in lattice QCD.
As above, an improved NRQCD action is used for the $b$ quark. 
The two light quarks are included with the ASQtad improved staggered action~\cite{Lepage:1998vj}
and the $s$ quark with the HISQ action.

The European Twisted Mass (ETM) Collaboration~\cite{Carrasco:2014cwa} presented a calculation 
which includes in the sea, besides two light mass degenerate quarks, also the dynamical strange and charm quarks 
with masses close to their physical values. 
The simulations were utilizing Wilson quarks with chirally twisted mass terms as pioneered 
by the ALPHA Collaboration~\cite{Frezzotti:2000nk,Frezzotti:2003xj}.
In this work, values for all first and second generation quark masses are computed.

The ALPHA Collaboration~\cite{Heitger:2013oaa} presented their own computation of $m_c$ 
with non-perturbatively ${\cal O}(a)$ improved Wilson quarks, at two lattice spacings $a$. 

The PACS--CS Collaboration~\cite{Namekawa:2011wt} investigated the charm quark system 
using the relativistic heavy quark action~\cite{Aoki:2001ra} for dynamical up-down and strange quark masses set to the physical values
in the Wilson-clover quark formalism.
The charm quark mass is determined from the spin-averaged mass of the $1S$ charmonium state, 

Reference~\cite{Laschka:2011zr} considers the heavy quark-antiquark potential
and matches the short-distance perturbative part to long-distance lattice QCD at an intermediate scale.
For the latter, QCD simulations with Wilson sea quarks by the SESAM and T$\chi$L Collaborations~\cite{Bali:2000vr} were employed.
The static and $1/m_b$ potentials are included, and values for $m_b$ and $m_c$ are obtained
from the empirical $\Upsilon(2S)$ and $h_c(1P)$ energies, respectively.

\subsection{$m_b$ and $m_c$ from QCD Sum Rules}
We next turn to sum rule analyses~\cite{Novikov:1977dq}.
Here the electronic partial widths of the narrow resonances are combined with threshold and continuum cross section data 
of $R_q \equiv \sigma(e^+e^-\to q\bar q)/\sigma(e^+e^-\to \mu^+\mu^-)$. 
What is actually measured is $R \equiv \sigma(e^+e^-\to {\rm hadrons})/\sigma(e^+e^-\to \mu^+\mu^-)$, 
and to find $R_c$ ($R_b$) one has to remove the contributions from the three (four) lighter quarks from $R$,
either theoretically by using the corresponding perturbative expressions, or by a form of sideband subtraction.
The theoretical moments ${\cal M}_n$ have been calculated to four-loop order corresponding to ${\cal O}(\alpha_s^3)$ in perturbative QCD,
where the first moment result appeared first~\cite{Chetyrkin:2006xg,Boughezal:2006px}
followed by $n=2$~\cite{Maier:2008he} and $n=3$~\cite{Maier:2009fz}.
Fairly precise numerical estimates for the higher moments were constructed using the low-energy, threshold, and high-energy
behavior of the vector-current correlator together with analyticity and Pad\'e approximations~\cite{Kiyo:2009gb}

Reference~\cite{Penin:2014zaa} uses the ${\cal O}(\alpha_s^3)$ approximation of the heavy-quark vacuum polarization function 
in the threshold region to determine $m_b$ from non-relativistic $\Upsilon$ sum rules. 
Relatively high moments are considered, where the quoted $m_b$ is extracted from ${\cal M}_{15}$. 
A correction of $- 25\pm 5$~MeV is applied at the end to account for $m_c \neq 0$~\cite{Hoang:1999us,Hoang:2000fm}.

Reference~\cite{Lucha:2013gta} uses the Borel QCD sum rule (featuring an exponential suppression towards large $s$)
for heavy-light pseudoscalar currents to obtain $m_b$ from the B-meson decay constant $f_B$, 
which in turn is taken as the average of precise lattice QCD determinations.
The correlator function is evaluated at three-loop order $\alpha_s^2$.

Reference~\cite{Hoang:2012us} determines $m_b$ from non-relativistic ($n \gg 1$) $\Upsilon$ sum rules 
with renormalization group improvement at next-to-next-to-leading logarithmic order. 
The theoretical moments are expanded simultaneously in $\alpha_s$ and $1/\sqrt{n}$
and the authors account for the summation of powers of the Coulomb singularities 
as well as of logarithmic terms proportional to powers of $\alpha_s \ln(n)$. 
The final result, $\overline{m}_b(\overline{m}_b) = 4235 \pm 55$~MeV, extracted from ${\cal M}_{10}$ is somewhat high,
but the aforementioned charm mass corrections~\cite{Hoang:1999us,Hoang:2000fm}, appropriate for large $n$, has not been applied. 

References~\cite{Bodenstein:2011fv,Bodenstein:2011ma} use finite energy QCD sum rules.
The analysis for $m_b$~\cite{Bodenstein:2011fv} employs integration kernels with arbitrary $n$,
but the eventually chosen kernel does not contain moments with $n < 0$ (${\cal M}_{0}$ does enter).
Note, that the quoted theoretical error of only $\pm 3$~MeV is entirely from perturbative QCD 
(estimated from the variation of the renormalization scale).
For $m_c$~\cite{Bodenstein:2011ma} the chosen sum rule is defined by the integration kernel $1 - s_0^2/s^2$,
where $s_0$ is the onset of perturbative QCD.
In this case, the quoted theory error of $\pm 4$~MeV is mostly from the variation of $s_0$.

The values obtained by Narison~\cite{Narison:2011rn} rely mostly on moments with $Q^2 \neq 0$~\cite{Narison:2010cg,Narison:2011xe},
but the result for $m_b$ includes a Borel sum rule analysis, as well.
The gluon condensates, $\langle \alpha_s G^2 \rangle$ and $\langle g^3 f_{abc} G^3 \rangle$, as well as the quark masses 
are extracted from different $Q^2$-dependent moments and their ratios.
Perturbative corrections are included up to ${\cal O}(\alpha_s^3)$ and non-perturbative terms up to order $\langle G^4 \rangle$. 
The central value for $m_c$ is strongly dominated by t${\cal M}_{15}$ with $Q^2 = 8 m_c^2$, 
and its uncertainty arises almost entirely from an uncertain non-relativistic Coulomb correction of $\pm 16$~MeV. 
The most precise result for $m_b$ is $\overline{m}_b(\overline{m}_b) = 4171 \pm 14$~MeV
from the ratio of ${\cal M}_{16}$ and ${\cal M}_{17}$ with $Q^2 = 8 m_b^2$, where the error is predominantly from perturbative QCD. 
The result from the Borel sum rule, $\overline{m}_b(\overline{m}_b) = 4212 \pm 32$~MeV, 
is slightly higher but consistent with Reference~\cite{Lucha:2013gta}.
    
Reference~\cite{Dehnadi:2011gc} determines $m_c$ from ${\cal O}(\alpha_s^3)$ perturbation theory.
A data clustering method is used to combine hadronic cross section data sets from different measurements.
The perturbative error dominates the overall uncertainty.  
  
References~\cite{Chetyrkin:2010ic,Chetyrkin:2009fv} present results for $m_b$ and $m_c$ using ${\cal M}_2$ and ${\cal M}_1$,
respectively.
The quoted theoretical errors are purely perturbative and amount to $\pm 3$ and $\pm 4$~MeV,
while the dominant errors are experimental and parametric.

\subsection{Other approaches to $m_b$ and $m_c$}
Reference~\cite{Ayala:2014yxa} determines $m_b$ using the NNNLO perturbative expression for the $\Upsilon(1S)$ mass.
This approach exploits the renormalon cancellation between the pole mass and the static potential.
Charm quark effects have been investigated and found to be negligible. 
The perturbative uncertainty dominates. 

Reference~\cite{Alekhin:2013nda} present a global fit of parton distributions at NNLO. 
The fit is based on the world data for deep-inelastic scattering (DIS), fixed-target data for the Drell-Yan process and includes
LHC data for the Drell-Yan process and the hadro-production of top-quark pairs. 
The output set includes $\alpha_s$, $m_c$, $m_t$ as well as higher twist (HT) terms.

Reference~\cite{Abramowicz:1900rp} is a combination of measurements of open charm production cross sections in 
$ep$-DIS at HERA in the kinematic range, $2.5 \mbox{ GeV}^2 < Q^2 < 2,000\mbox{ GeV}^2$,
and Bjorken scaling variable $0.00003 < x < 0.05$. 
These charm data together with the combined inclusive DIS cross sections from HERA 
are used for a detailed NLO QCD analysis.

\section{Strong Coupling Constant}
The most recent precision determinations of $\alpha_s(M_Z)$ are listed in Table~\ref{tab:as}.
Except for the result returned by the precision electroweak fit, all uncertainties are strongly theory dominated. 
In those cases in which the original references quote asymmetric error bars they have been symmetrized for the table. 
The level of agreement is certainly worse than for $m_b$ and $m_c$.
In particular, the two result from $e^+ e^-$ thrust variables are significantly lower than those from the lattice. 
The weighted average is
\begin{equation}
\label{eqas}
\alpha_s(M_Z) = 0.11743 \pm 0.00035
\qquad\qquad(\mbox{weighted uncorrelated average)}.
\end{equation}
Even ignoring possible correlations, the $\chi^2$-value of 34.9 for 12~effective degrees of freedom 
is very poor, where the probability for a larger $\chi^2$ is less than 0.05\%.
This can be explained at least in part by underestimates or neglect of some error components.
For example, the PDF + HT fit does not include an error from unknown higher perturbative orders
and is the greatest contributor to the overall $\chi^2$.
Inflating all errors by a factor of about 1.7 would result in a more reasonable $\chi^2 \approx 12$,
corresponding to a combined error of $\pm 0.0006$.

\begin{table}
\begin{tabular}{clllcc}
\hline
     \tablehead{1}{c}{b}{Result}
  & \tablehead{1}{c}{b}{Approach}
  & \tablehead{1}{c}{b}{Observable}
  & \tablehead{1}{c}{b}{Authors/Group}
  & \tablehead{1}{c}{b}{Reference} 
  & \tablehead{1}{c}{b}{arXiv.org} \\
\hline
$0.11856 \pm 0.00053$ & lattice ($n_f = 4$) & pseudoscalar current & HPQCD & \cite{Chakraborty:2014aca} & 1408.4169 \\
$0.1166\ph{0} \pm 0.0010\ph{3}$ & lattice ($n_f = 2 + 1$) & static potential & Bazavov \etal & \cite{Bazavov:2014soa} & 1407.8437 \\
$0.1165\ph{0} \pm 0.0039\ph{3}$ & NLO fit & jet cross sections & H1 & \cite{Andreev:2014wwa} & 1406.4709 \\
$0.1192\ph{0} \pm 0.0027\ph{3}$ & global fit & precision electroweak & Erler & \cite{Moch:2014tta} & 1405.4781 \\
$0.1196\ph{0} \pm 0.0011\ph{3}$ & lattice ($n_f = 2 + 1 + 1$) & ghost-gluon vertex & ETM & \cite{Blossier:2013ioa} & 1310.3763 \\
$0.1132\ph{0} \pm 0.0011\ph{3}$ & PDF + HT fit & DIS & Alekhin \etal & \cite{Alekhin:2013nda} & 1310.3059 \\
$0.1151\ph{0} \pm 0.0028\ph{3}$ & NNPDF fit & $t\bar{t}$ cross section & CMS & \cite{Chatrchyan:2013haa} & 1307.1907 \\
$0.1174\ph{0} \pm 0.0014\ph{3}$ & RGOPT & $f_\pi$ & Kneur and Neveu & \cite{Kneur:2013coa} & 1305.6910 \\
$0.1184\ph{0} \pm 0.0020\ph{3}$ & BRGSPT & $\tau$ decays & Abbas \etal & \cite{Abbas:2012fi} & 1211.4316 \\
$0.1131\ph{0} \pm 0.0025\ph{3}$ & NNLO fit & $e^+ e^-$ thrust & Gehrmann \etal & \cite{Gehrmann:2012sc} & 1210.6945 \\
$0.1140\ph{0} \pm 0.0015\ph{3}$ & SCET & $e^+ e^-$ thrust & Abbate \etal & \cite{Abbate:2012jh} & 1204.5746 \\
$0.1191\ph{0} \pm 0.0022\ph{3}$ & FOPT & $\tau$ decays & Boito \etal & \cite{Boito:2012cr} & 1203.3146 \\
$0.1201\ph{0} \pm 0.0030\ph{3}$ & NNLO fit & $e^+ e^-$ event shapes & OPAL & \cite{OPAL:2011aa} & 1101.1470 \\
\hline
\end{tabular}
\caption{Most recent determinations of the strong coupling constant $\alpha_s^{(n_f = 5)}(M_Z)$ in the five-flavor effective theory.
Only results with a quoted uncertainty not exceeding $\pm 0.004$ are included.
Not considered are quenched lattice simulations or those with less than three dynamical flavors.
See body of text for details.}
\label{tab:as}
\end{table}

\subsection{$\alpha_s$ from Lattice Gauge Theory}
There are three very precise lattice results, all released within the last year. 
The most precise one is from the analysis of the pseudoscalar current in Reference~\cite{Chakraborty:2014aca}, 
which was already remarked on in the context of $m_b$ and $m_c$.

The result $\alpha_s(M_Z) = 0.1166^{+0.0012}_{-0.0008}$ obtained from the static quark-antiquark energy 
in Reference~\cite{Bazavov:2014soa} is somewhat lower and includes a comprehensive and detailed estimate of the error sources. 
Quarks are treated using the HISQ action procured by the HotQCD Collaboration~\cite{Bazavov:2014pvz}.
The strange quark mass was fixed to its physical value and the light quark masses were taken as degenerate. 

On the other hand, the result by the ETM Collaboration~\cite{Blossier:2013ioa} is near the upper end of the values of Table~\ref{tab:as}.
It is obtained from simulations where the ghost and gluon propagators are computed from gauge configurations 
based on $n_f = 2 + 1 + 1$ twisted-mass lattice flavors~\cite{Frezzotti:2000nk}.
The strong coupling is converted from the ghost-gluon coupling renormalized in the MOM Taylor scheme~\cite{Taylor:1971ff}.
  
\subsection{$\alpha_s$ from weak decays}
In the global electroweak fit~\cite{Erler:2014}, $\alpha_s$ is predominantly constrained by 
the total $Z$ boson decay width, the hadronic peak cross section $\sigma^0_{\rm had}$ of the $Z$ lineshape, 
and the leptonic $Z$ branching ratios, but other measurements, SM parameters, 
and new physics generally enter into the picture~\cite{Moch:2014tta,Erler:2011ux}. 
The experimental correlations between the observables are small, known, and included.
The parametric uncertainties introduced by the imperfectly known SM parameters are in general non-Gaussian,
in particular the one from the weak mixing angle, but this can be treated exactly in fits.
The perturbative series for massless quarks is fully known and included up to ${\cal O}(\alpha_s^4)$~\cite{Baikov:2008jh}.
The ${\cal O}(\alpha_s^4)$ axial-vector singlet piece involving top quark loops previously dominated the theory error~\cite{Erler:2011ux},
but the terms without further $M_Z^2/m_t^2$ suppressions are now known~\cite{Baikov:2012er},
with other unknown corrections affecting $\alpha_s$ by $< 10^{-4}$.
Note that $\sigma^0_{\rm had}$ deviates by 1.7~standard deviations from the SM which drags down average.
The fit result, $\alpha_s = 0.1192 \pm 0.0027$, is rather insensitive to new physics corrections affecting only the weak
gauge boson propagators (oblique corrections), but if one allows special new physics corrections to the $Zb\bar{b}$-vertex
the electroweak fit returns a much smaller value, $\alpha_s = 0.1167 \pm 0.0038$.

$\alpha_s$ can also be extracted from $\tau$ lepton decays, 
where perturbation theory is known to ${\cal O}(\alpha_s^4)$~\cite{Baikov:2008jh}, as well.
For example, for non-strange decays one can write in fixed-order perturbation theory (FOPT)
\begin{equation}
\label{atau}
\Gamma_{ud}^{\rm theo} = {G_F^2 |V_{ud}|^2 m_\tau^5 \over 64 \pi^3} S(m_\tau, M_Z) \left(1 + {3\over 5} {m_\tau^2\over M_Z^2} \right)
\left(1 + {\alpha_s(m_\tau)\over \pi} + 5.202  {\alpha_s^2\over\pi^2} + 26.37 {\alpha_s^3\over\pi^3} + 127.1 {\alpha_s^4\over\pi^4} - 
1.393 {\alpha\over\pi} + \delta_q \right)
\end{equation}
where $G_F |V_{ud}|$ is the effective Fermi constant, 
$S(m_\tau, M_Z) = 1.01907 \pm 0.0003$ is a logarithmically enhanced electroweak correction factor 
with higher orders re-summed~\cite{Erler:2002mv}, 
and $\delta_q$ contains higher dimensional terms in the operator product expansion, as well as duality violating effects.
However, there is a controversy whether higher order terms should be re-summed~\cite{Le Diberder:1992te}
in what is called contour-improved perturbation theory (CIPT), 
or whether one should strictly apply FOPT as argued in Reference~\cite{Beneke:2008ad}.
CIPT amounts to a re-organization of perturbation theory in terms of functions $A_n(\alpha_s)$ which only in the limit $\alpha_s \to 0$
approach the powers $\alpha_s^n$ of FOPT.
The coefficients of the $A_n(\alpha_s)$ coincide with the expansion coefficients of the Adler function.
The problem is that CIPT yields significantly larger values of $\alpha_s$ than FOPT.
This is puzzling, as in either approach one seemingly obtains a reasonably convergent series for the known terms,
but they appear not to converge towards each other.
Table~\ref{tab:as} contains in addition to the FOPT value of Reference~\cite{Boito:2012cr},
an alternative analysis based on Borel and renormalization group summed perturbation theory (BRGSPT)~\cite{Abbas:2012fi}
which combines renormalization-group invariance (a feature shared by CIPT) with knowledge about the large-order behavior of the series. 
The authors note remarkable convergence properties up to high orders. 
Numerically this approach yields values that are close to (and even slightly lower than) those of FOPT.
$\delta_q$ can be constrained from fits to $\tau$ spectral functions.
Again, there are numerically non-negligible differences in the details of various evaluations,
with Reference~\cite{Davier:2008sk} giving higher values of $\alpha_s$ than Reference~\cite{Boito:2012cr}.

The experimental information~\cite{Agashe:2014kda} derives from the leptonic branching ratios, 
$\tau[{\cal B}_\ell] = \hbar {\cal B}_\ell^{\rm expt}/\Gamma_\ell^{\rm theo}$,
as well as the direct $\tau$ lifetime measurements, $\tau_{\rm direct}^{\rm expt}$.
An update of the analysis in Reference~\cite{Erler:2014} is given here:
\begin{eqnarray}
{\cal B}_e^{\rm expt} = 0.1783 \pm 0.0004 \Longrightarrow \tau [{\cal B}_e^{\rm expt}] &=& 291.15 \pm 0.65 \mbox{ fs} \\
{\cal B}_\mu^{\rm expt} = 0.1741 \pm 0.0004 \Longrightarrow \tau [{\cal B}_\mu^{\rm expt}] &=& 291.85 \pm 0.67 \mbox{ fs} \\ \hline
{\cal B}_{e,\mu}^{\rm expt} (\rho_{e\mu} = 0.13)  \Longrightarrow \tau [{\cal B}_{e,\mu}^{\rm expt}] &=& 291.49 \pm 0.50 \mbox{ fs} \\ 
\tau_{\rm direct}^{\rm expt} &=& 290.3 \pm 0.5 
\mbox{ fs\quad\quad(PDG~\cite{Agashe:2014kda} incl. new Belle result~\cite{Belous:2013dba})} \\ \hline
\tau_{\rm direct}^{\rm expt} \equiv\tau [{\cal B}_{e,\mu}^{\rm expt},\tau_{\rm direct}^{\rm expt}] &=& 290.90 \pm 0.35 \mbox{ fs}  \\
R \equiv \Gamma_{ud}/\Gamma_e &=& 3.479 \pm 0.007 \quad\quad (\delta_{\rm QCD} = 0.1977 \pm 0.0025)  \\
\Longrightarrow \alpha_s(M_Z)[\tau_\tau] &=& 0.1195^{+0.0022}_{-0.0020} \quad\quad \mbox{ (september 2014 using FOPT)}
\end{eqnarray}
Here $\rho_{e\mu}$ is the experimental correlation coefficient between
 ${\cal B}_e^{\rm expt}$ and ${\cal B}_\mu^{\rm expt}$~\cite{Agashe:2014kda}.
The PQCD error is taken as the size of the ${\cal O}(\alpha_s^4)$-term in Equation~\eqref{atau},
which covers almost the entire range from CIPT to an assumed continuation of the approximately geometric series in Equation~\eqref{atau}.

\subsection{Other approaches to $\alpha_s$}
The H1 Collaboration at HERA~\cite{Andreev:2014wwa} presents measurements of inclusive jet, dijet and trijet differential cross sections 
in neutral current DIS for boson virtualities $150 \mbox{ GeV}^2 < Q^2 < 15,000\mbox{ GeV}^2$. 
Normalized double differential jet cross sections are also measured.
Comparison to perturbative QCD calculations in NLO are used to determine $\alpha_s$.

CMS~\cite{Chatrchyan:2013haa} compares a measurement of the inclusive $t\bar{t}$ production cross section
to the NNLO QCD prediction with various PDFs to determine $m_t$ or $\alpha_s$. 
This is the first determination of $\alpha_s$ using events from top-quark production.

Reference~\cite{Kneur:2013coa} uses renormalization group optimized perturbation theory (RGOPT)
to calculate the ratio of the pion decay constant $f_\pi$ to the QCD scale $\Lambda_{\rm QCD}$. 
Using the experimental $f_\pi$ as input this provides a determination of 
$\alpha_s (M_Z) = 0.1174^{+.0010}_{-.0005} \pm 0.001_{f_0} \pm 0.0005_{\rm evol}$,
where the first error is the theoretical uncertainty of the method,
the second comes from the present uncertainty of $f_0$ ($f_\pi$ in the exact chiral limit),
and the last is from the evolution to $M_Z$.

Two of the values of $\alpha_s$ in Table~\ref{tab:as} are extracted from thrust distributions in $e^+ e^-$ annihilation.
Reference~\cite{Gehrmann:2012sc} performs fits using a NNLL + NNLO perturbative description including $m_b$ effects to NLO.
The fits of Reference~\cite{Abbate:2012jh} employ cumulant moments and use predictions of the full spectrum for thrust 
including ${\cal O}(\alpha_s^3)$ fixed order results, resummation of singular N$^3$LL contributions, 
and a class of leading power corrections in a renormalon-free scheme. 
The authors see this as a cross-check of their earlier, more precise analysis~\cite{Abbate:2010xh}, 
which gave $\alpha_s (M_Z) = 0.1135 \pm 0.0010$.

The OPAL Collaboration~\cite{OPAL:2011aa} combined all of its measured hadronic event shape variables from $e^+ e^-$ annihilation 
to determine $\alpha_s$. 
The result in Table~\ref{tab:as} is based on QCD predictions complete to NNLO.

\section{Summary}
The determinations of the three heavy quark masses and of the strong coupling constant have all reached a relative precision of about 0.5\%.
New and precise results for $m_t$ by D\O\ and from 8~TeV data by CMS --- both in the seminal lepton + jets channel --- 
are in good agreement with their respective previous results but there is a conflict of at least $2.7~\sigma$ between them.
As for $m_b$ and $m_c$, all the results (with one exception in each case) are consistent with
\begin{eqnarray}
\bar m_b(\bar m_b) = 4175 \pm 20 \mbox{ MeV \qquad\qquad 
(PDG~\cite{Agashe:2014kda}: } & \bar m_b(\bar m_b) = 4180 \pm  30~{\rm MeV}) \\
\bar m_c(\bar m_c) = 1276 \pm 6 \mbox{ MeV \qquad\qquad 
(PDG~\cite{Agashe:2014kda}: } & \bar m_c(\bar m_c) = 1275 \pm  25~{\rm MeV}) \\
\alpha_s(M_Z) =  0.1174 \pm 0.0006  \mbox{ \qquad\qquad
(PDG~\cite{Agashe:2014kda}: } &  \alpha_s(M_Z) =  0.1185 \pm 0.0006)
\end{eqnarray}
but there is wider scatter in the $\alpha_s(M_Z)$ values with its average dragged down by PDF fits and thrust variables.

In the future, it will not only be important to reduce the individual and overall uncertainties, 
but also to develop more rigorous methods for error estimations.
One step in this direction was taken in Reference~\cite{Erler:2002bu},
where the continuum region of $R_q$ has been derived theoretically from sum rules and then compared with experimental data.  
In this way the uncertainty in $m_b$ and $m_c$ is self-calibrating~\cite{Erler:2014a}.

\begin{theacknowledgments}
It is a pleasure to thank the organizers for the invitation and for ensuring an exciting and memorable conference. 
This work was supported by PAPIIT (DGAPA--UNAM) project IN106913 and
CONACyT (M\'exico) project 151234.
\end{theacknowledgments}

\bibliographystyle{aipproc}   

\end{document}